\documentstyle[11pt,moriond,epsf]{article}
\def \Et {{\rm E}_{\rm T}}
\def \Pt {{\rm P}_{\rm T}}

\def\Z0{${\em Z^0\/}$}

\def\r#1 {$^{#1}$}

\newcommand{\et}{{\rm E}_{\scriptscriptstyle\rm T}}

\newcommand{\met}{\mbox{$\protect \raisebox{.3ex}{$\not$}\et \ $}}

\newcommand{\ppbar}{p\bar{p}}

\newcommand{\ttbar}{t\bar{t}}

\def \mtop {$M_{top} \ $}

\def \tljx {\ttbar \rightarrow l \nu q \bar{q} b \bar{b} X}

\newcommand{\gev}  { {\rm GeV}}
\newcommand{\gevc} { {\rm GeV/c  }}
\newcommand{\gevcc}{ {\rm GeV/c^2}}

\def\gepsfcentered#1{
  \def\testit{#1}
  \def\lbracket{[}
  \ifx\testit\lbracket
    \let\dofilecmd=\gepsfwithopt
  \else
    \let\dofilecmd=\gepsfnoopt
  \fi
  \dofilecmd}

\def\gepsfnoopt#1{
  \begin{center}
  \leavevmode
  \epsffile{#1}
  \end{center}}

\def\gepsfwithopt#1 #2 #3 #4]#5{
  \begin{center}
  \leavevmode
  \gepsfmaxx=0.94\textwidth
  \epsffile[#1 #2 #3 #4]{#5}
  \end{center}}

\newdimen\gepsfmaxx
\def\epsfsize#1#2{
  \ifnum \epsfxsize=0
    \ifnum \epsfysize=0
      \ifnum #1 > \gepsfmaxx
        \gepsfmaxx
      \else
        #1
      \fi
    \else
      \epsfxsize
    \fi
  \else
    \epsfxsize
  \fi
}

\def\be{\begin{equation}}
\def\ee{\end{equation}}
\def\bea{\begin{eqnarray}}
\def\eea{\end{eqnarray}}
\begin{document}
\vspace*{4cm}
\title{MEASUREMENT OF THE TOP QUARK MASS}

\author{ STEVEN R. BLUSK }

\address{Department of Physics and Astronomy, \\
University of Rochester, \\ 
Rochester, NY, 14627 \\
\vskip0.25in
\it{for the}
\vskip0.25in
\bf{CDF and D0 Collaborations}}

\maketitle\abstracts{
  The first evidence\cite{cdf-evidence} and subsequent 
discovery\cite{cdf-discovery,d0-discovery} of the top quark
was reported nearly 4 years ago. Since then, CDF and D0
have analyzed their full Run 1 data samples, and analysis techniques
have been refined to make optimal use of the information.
In this paper, we report on the most recent measurements of the top quark mass,
performed by the CDF and D0 collaborations at the Fermilab Tevatron.
The CDF collaboration has performed measurements of the top quark mass
in three decay channels from which the top quark mass is measured to
be $175.6\pm 6.8\ \gevcc$. The D0 collaboration
combines measurements from two decay channels to obtain a top quark
mass of $172.1\pm 7.1\ \gevcc$. Combining the measurements from the
two experiments, assuming a 2 $\gevcc$ correlated systematic
uncertainty, the measurement of the top quark mass at the Tevatron is
$173.9\pm 5.2\ \gevcc$. This report presents the measurements of the 
top quark mass from each of the decay channels which contribute to 
this measurement.}

\newpage
\section{Introduction}
  The top quark is defined as the $I_{3}=+1/2$ member of a
weak SU(2) isodoublet that also contains the $b$ quark. In 
$\ppbar$ collisions, top quarks
are produced primarily in $\ttbar$ pairs and are expected to 
decay through the electroweak interaction to a W-boson and 
b-quark. In the standard 
model, the branching fraction for $t\to Wb$ is expected to be nearly 
100\%. The decay width~\cite{twidth} is calculated to
be $1.6-1.7 \ \gev$ for masses between $150-180 \ \gevcc$.
The top quark mass is sufficiently large that top-flavored hadrons are
not expected to form~\cite{notophad}. 

  The top quark mass, \mtop, is an important parameter in 
calculations of electroweak processes since its mass is approximately
35 times larger than that of the next heaviest fermion. Like other fermions, 
its mass is taken to be an unpredicted parameter of the standard 
model~\cite{SM}. Calculations of electroweak radiative corrections
relate the top quark and the W-boson ($M_W$) masses to that of the
Higgs boson. Precision measurements of \mtop and $M_W$ can therefore
aid in searches for the Higgs boson. 

\section{Overview of Top Quark Mass Measurement}

   In $\sqrt{s}=1.8$ TeV $\ppbar$ collisions, top quarks are expected
to be produced in pairs mainly through the subprocess $q\bar{q}\to \ttbar$.
At the Tevatron, both CDF and D0 select samples of events which are
consistent with containing a $\ttbar$ pair. Since each W-boson may
decay leptonically or hadronically, this leads to three categories
of events. Events which have 2 high $\Pt$ isolated leptons 
(both W-bosons decay leptonically), missing transverse energy ($\met$) 
\cite{metdef} and two or more jets (from the b-quarks) are classified 
as {\it{Dilepton}} events. Events which have one isolated high $\Pt$ lepton, $\met$,
and 4 or more jets are referred to as {\it{l+jets}} events. Events containing
6 or more jets and no high $\Pt$ isolated leptons are classified as
{\it{All-hadronic}} events. Neither CDF nor D0 try to reconstruct the
top mass from candidate $\ttbar\to\tau+X$ events. Ignoring final states
containing $\tau$'s, the branching ratio for $\ttbar$ into these three 
decay channels is 5\%, 30\%, and 44\% respectively.

    Both CDF and D0 have measured the top quark mass in the
Dilepton  and l+jets channels. CDF has also made
a measurement of the top quark mass in the All-Hadronic channel.
A general procedure is used for extracting the top quark mass.
For each decay channel, we choose a variable (or a set of 
variables) which can be measured in the data sample, and is highly 
correlated with \mtop. Examples of such variables
include event-by-event top mass, jet $\Et$, $\met$, total scalar
$\Et$, etc. The distribution of a given variable
is modelled using Monte Carlo (MC) simulations of $\ttbar$ signal 
and background. The $\ttbar$ signal distributions are modelled
using the {\small{HERWIG}}\cite{herwig} MC simulation and are produced for
a wide range of input top quark masses, ranging from $\approx$120 
to 220 $\gevcc$. Background distributions are derived from a 
combination of Monte Carlo and/or data, depending on the channel.
The modelled signal and background distributions are generically
referred to as {\it{templates}}. 
The top quark mass is extracted from a maximum likelihood procedure 
which compares the observed data distribution(s) to the $\ttbar$ signal 
and background templates. Assumptions made in the modelling of
signal and background are included as sources of systematic uncertainty.

\section{Top Quark Mass Measurement in the Dilepton Topology}

   Both CDF\cite{cdf-dilepton} and D0\cite{d0-dilepton} have reported 
measurements of the top quark mass in the Dilepton decay topology.
Because these events are presumed to have two neutrinos, the 
kinematics of the $\ttbar$ system (including the top quark mass)
cannot be uniquely determined. A measurement of the top quark mass can
however be made by comparing the measured observables with the
expectation from $\ttbar$ and background. 
Each experiment employs two techniques to measure the top quark
mass from their respective samples.

\subsection{CDF Dilepton Mass Measurements}

    The first Dilepton mass measurement uses the
energies of the 2 highest $\Et$ jets in the candidate events
as a discriminator for the top quark mass measurement. 
From a sample of 8 events, with an expected background of
1.1$\pm$0.3 events, the analysis yields a top quark mass of 
159$\pm$23(stat)$\pm$11(syst) $\gevcc$. A second analysis
exploits the fact that, in the W-boson rest frame of the $t\to Wb$
decay, the top quark mass can be related to the invariant mass of 
the b-jet and lepton ($M_{lb}$) to which it decays. Among the
two possible pairings of the 2 leptons with the 2 highest $\Et$
jets, the pair with the smaller sum of invariant masses is
chosen. From this analysis, the top quark mass is measured to
be 163$\pm$20(stat)$\pm$6(syst) $\gevcc$. The jet energy scale,
and signal and background modelling dominate the systematic
uncertainty. The two measurements are combined taking into 
account the correlations in uncertainties to obtain a top quark 
mass of 161$\pm$17(stat)$\pm$10(syst) $\gevcc$. 

\subsection{D0 Dilepton Mass Measurements}

Both techniques which D0 employs uses MC simulations to calculate
a probability that the observed kinematics of the event are consistent 
with top of an assumed mass. This probability is mapped out as a function
of assumed top mass in the range from 80 to 280 $\gevcc$.
The probability in the first technique is based on: (a) the probability for the
charged leptons to have come from $\ttbar$ decay at the assumed mass,
and (b) the probability for the event to have been produced by valence 
quarks with momentum fractions $f(x)$ and $f(\bar{x})$. 
The second technique calculates a probability based on the difference
between the measured and calculated $\met$ in the event.
The calculation of the $\met$ is facilitated by assuming values
for the pseudorapidity of the two neutrinos ($\eta_1$ and $\eta_2$),
and then integrating over the $\eta_1$, $\eta_2$ phase space.
For both techniques, detector resolution effects are included as well as
trying both lepton-(b-quark) combinations.
Based on the data sample of 6 events and an expected background
of 1.42 events, the mass is measured to be 
168.1$\pm$12.4(stat)$\pm$3.6(syst) $\gevcc$ using the first technique,
and 169.9$\pm$14.8(stat)$\pm$3.6(syst) $\gevcc$ using the second technique.
The dominant systematics are the same as those mentioned for
the CDF dilepton analyses. The measurements are combined to
obtain a top quark mass of 168.4$\pm$12.3(stat)$\pm$3.6(syst) $\gevcc$. 

\section{Top Quark Mass Measurement in the l+jets Topology}

Both CDF\cite{cdf98-ljprl} and D0\cite{d0-ljprl} have measured 
the top quark mass in the l+jets topology. The l+jets decay channel 
currently allows for the
most accurate determination of the top quark mass. The 
measurement benefits from a relatively large branching ratio
and the ability to fully reconstruct the top mass on an event-by-event
basis. The momentum of all final state particles are assumed
to be measured, with the exception of the unobserved neutrino, $\nu$.
The transverse momentum components of the neutrino however
can be inferred from the $\met$ in the event, leaving only
its Z-component unmeasured. Three kinematic constraints are
implied by the $\tljx$ decay hypothesis. Namely, we require
that $M_{{\it l}\nu}=M_W$, $M_{jj}=M_W$, and $M_{{\it l}\nu b}=M_{jjb}$, 
where  $b$ = $b$ or $\bar{b}$ jet, $j$=non-$b$ jet, $M_W$=W-boson mass,
and $M_{xx(x)}$ are invariant masses formed from the reconstructed 
4-vectors of the indicated objects. Because $\ttbar$ events are expected 
to contain 2 b-quarks, as compared to only a few percent for background 
events, identifying b-jets improves the signal to background ratio (S/B). 
B-jets are tagged by either
reconstructing the secondary decay vertices of B-hadrons (SVX tagged jet), 
or by identifying charged tracks which are consistent with coming from
semileptonic b-hadron decay (SLT tagged jet). Each event is fit to the 
$\tljx$ topology, requiring b-tagged jets to be assigned to b-quarks, 
All allowable permutations are tried, and the top mass 
corresponding to the solution with the lowest $\chi^2<10$ is taken as 
the reconstructed top mass ($M_{rec}$) for a given event. The ensemble of 
reconstructed masses (one per event) are fit using a likelihood 
procedure to extract the best estimate of the top quark mass. 

\subsection{CDF l+jets Mass Measurement}
   
    The initial sample selection requires 1 isolated lepton 
with $\Pt>20 \ \gevc$, $|\eta |<1$, $\met>20 \ \gev$, 
$\ge 3$ jets with $\Et>15 \ \gev$, 
$|\eta |<2.0$, and a fourth jet with $\Et>8 \ \gev$ and $|\eta |<2.4$.
After applying these selection criteria and
the $\chi^2$ cut on the kinematic fit, the sample consists of 
151 events, of which $\approx$35\% are estimated to be from $\ttbar$. 
Motivated by the different S/B and top mass resolution for events with 
SVX, SLT, or no tags,
MC simulations were used to determine an optimal way to partition the 
sample as to obtain the best precision on the top quark mass
measurement\cite{kirsten}. Studies showed that an optimal partitioning
consists of subdividing the events into the four statistically
independent categories shown in Table~\ref{t-sample}. The table
shows the numbers of events in each subsample, the expected 
background fraction\cite{bgfrac} $x_b$, and the fitted top mass.

\begin{table}[ht]
\begin{center}
\begin{tabular}{|c|c|c|c|c|}
\hline
Data Sample                    & \#Events & $x_b$(\%)  & Top Mass ($\gevcc$) \\ 
\hline\hline
SVX Double (2 SVX tagged jets) &  5  & $5^{+4}_{-2}$   & $170.1\pm9.3$  \\ 
SVX Single (1 SVX tagged jet)  & 15  & $13^{+5}_{-4}$  & $178.0\pm7.9$ \\
SLT tagged ($\ge$1 SLT tagged jets, no SVX tags) & 14  & $40^{+9}_{-9}$        & $142.1^{+33}_{-14}$ \\ 
No Tags ($\ge$4 jets with $\Et>15\ \gev$, $|\eta |<2.0$) 
& 42  & $56^{+14}_{-17}$ & $180.8\pm9.0$ \\ 
\hline
Combined    &                        76   &  -         & $175.9\pm 4.8$ \\  
\hline
\end{tabular}
\caption{Subsamples used in the top quark mass analysis,
expected background fractions, and the measured top
quark mass in each sample. The last line shows the
results from the combined likelihood fit to all four subsamples.}
\label{t-sample}
\end{center}
\end{table}

\begin{figure}[h]
\epsfxsize=6.0in
\hglue1truept\hspace{0.3in}
\epsffile{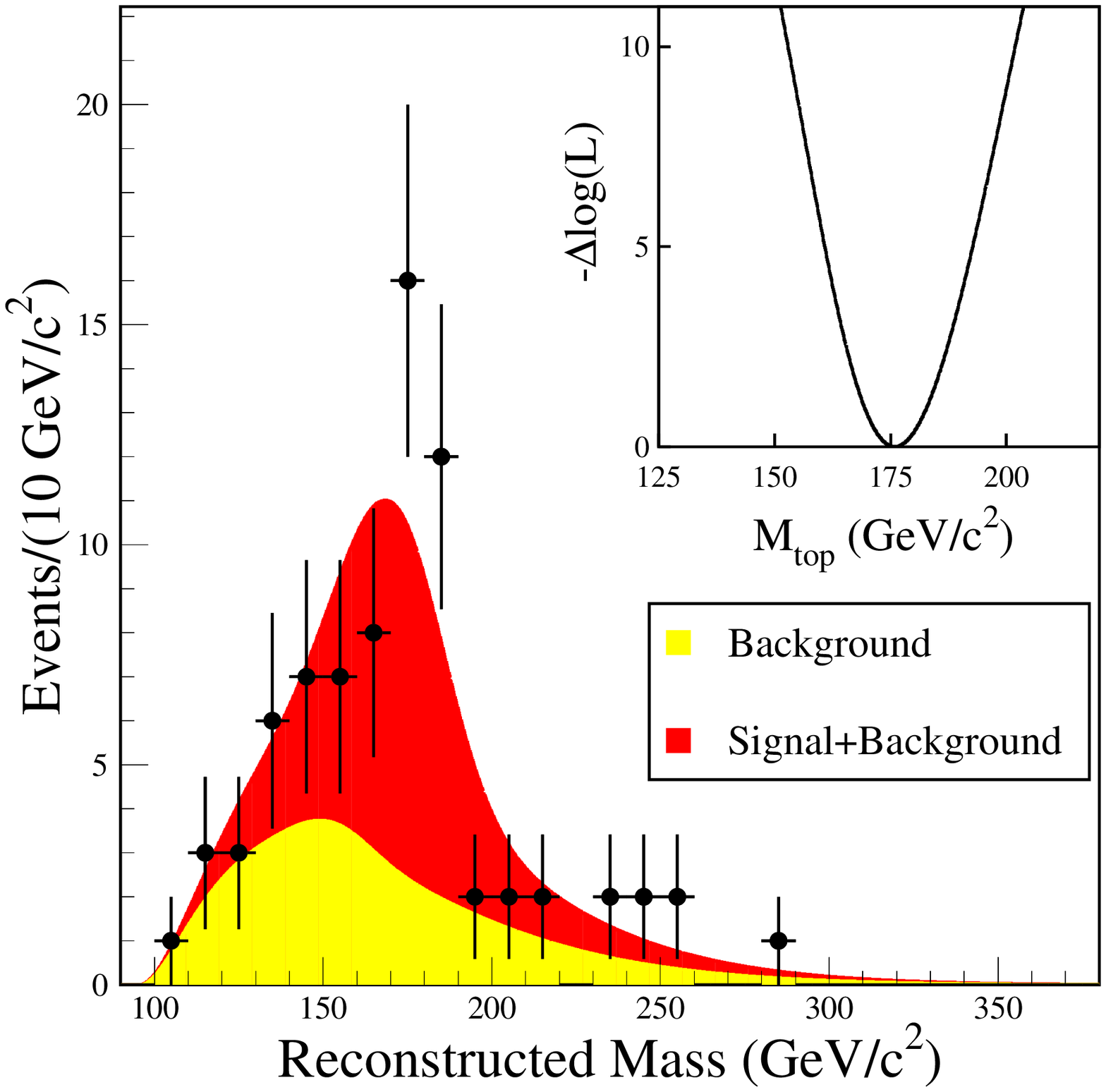}
\caption{The results of applying the likelihood procedure 
to the combined subsamples.
The figure shows the data (points), fitted background (light shaded region), 
and fitted signal plus background (darker shaded region). The inset shows 
the shape of $-\log{\cal{L}}$
versus the generated top mass from which we extract the best estimate of the top
mass and the statistical uncertainty.}
\label{opt1_err_mod}
\end{figure}


  Since the four subsamples are statistically independent, the joint
likelihood for the 76 events is given by the product of the four 
subsample likelihoods. Each subsample likelihood uses parameterized
signal and background probability densities, along with a 
constraint on the the background fraction (see Table~\ref{t-sample}), 
to evaluate the likelihood for observing the measured set of 
masses as a function of the top mass. The results of the likelihood fit
for each subsample and the combined result are shown in
Table~\ref{t-sample}. The results of the combined fit for the 76
data events and are shown in Fig.~1. Including the systematic
uncertainties presented in Table~\ref{t-ljetsys}, the top quark 
mass is measured to be 175.9$\pm$4.8(stat.)$\pm$4.9(syst.) $\gevcc$ 
for events in the l+jets topology.

\subsection{D0 l+jets Mass Measurement}

  The event selection for the D0 analysis is similar to CDF. The
main differences are a larger $|\eta |$ cut for the charged lepton (from
the W-boson decay) and 
all four jets are required to have $\Et>15 \ \gev$.
After the event selection criteria and the 
kinematic $\chi^2$ cut, 77 events remain of which 5 are SLT tagged (D0 did not have a silicon vertex 
detector in Run 1).  D0 extends the likelihood procedure by including an additional variable $D$ 
which provides discrimination
between signal and background events. Two versions of the discriminant
are tried. The first is a low-bias version $D_{lb}$, and the second,
$D_{NN}$, was the output from a neural network trained on $\ttbar$ signal
and background events. 
Both discriminants provide similar levels of 
discrimination between signal and background events. Two-dimensional signal 
templates of $D$ vs $M_{rec}$ are generated for many input values for the  
top mass and a single template for the background.
For each top mass value, a likelihood is minimized with respect to the
number of signal and background events. The corresponding negative log-likelihood
values are fit to a quadratic function of top mass, and the measured top
mass is taken to be where the function is a minimum. The fits to the data 
are presented in Table~\ref{t-d0mass} and shown in Fig.~2
For illustrative purposes, the 
data sample is separated into (a) signal depleted 
and (b) signal enhanced subsamples. Also shown are the likelihood 
fits for the two methods. Including the
systematic uncertainties shown in Table~\ref{t-ljetsys}, the top quark
mass is measured to be 173.3$\pm$5.6(stat.)$\pm$5.5(syst.) $\gevcc$. 

\begin{figure}
\epsfxsize=6.0in
\hglue1truept\hspace{0.0in}
\epsffile{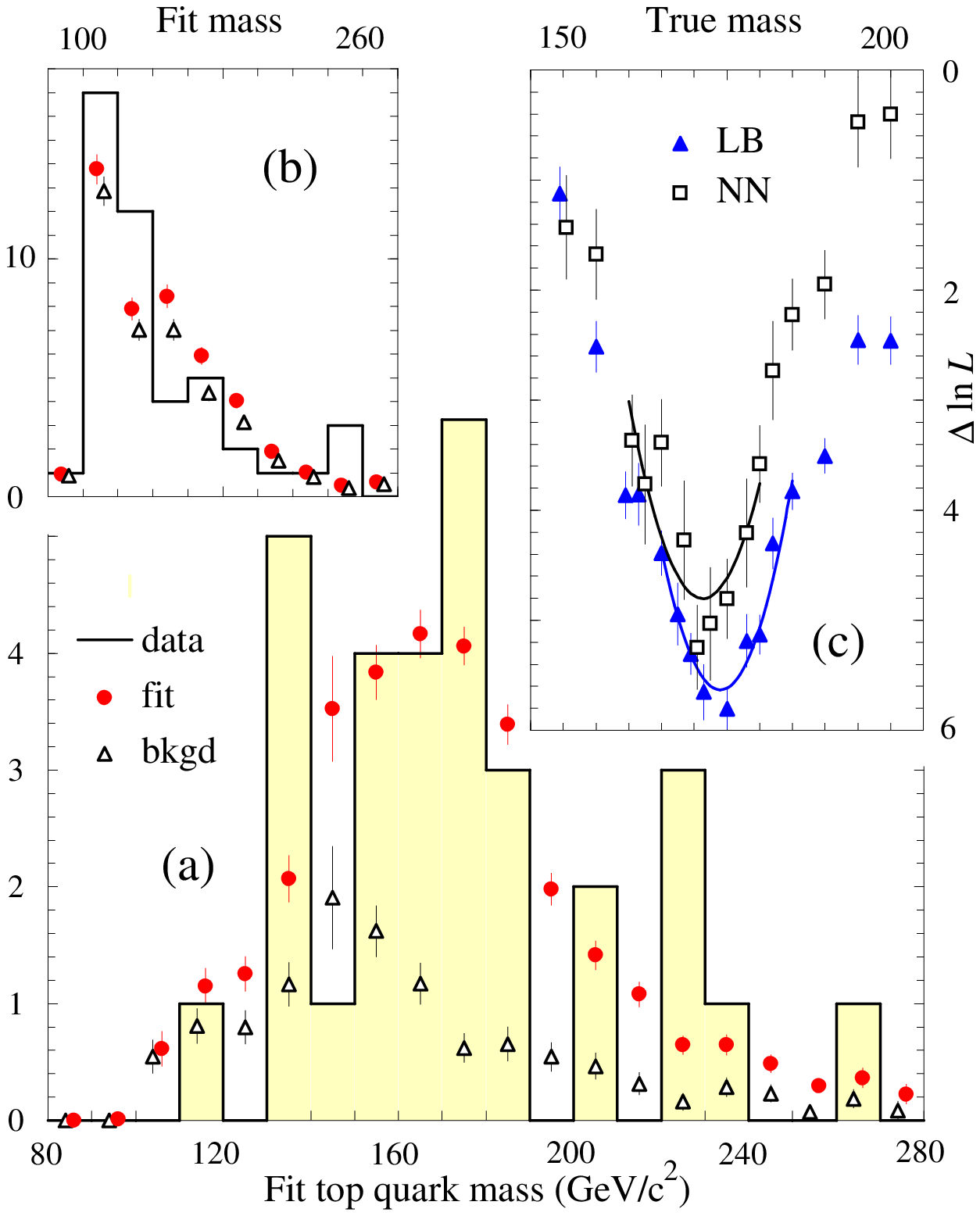}
\caption{The fit results from the LB and NN likelihood procedures
applied to the 77 event data sample. (a) Events which have $D_{lb}>0.43$
and (b) $D_{lb}<0.43$. The histogram shows the data, and the points are
the MC fit results. (c) shows the fit to $-log{\cal{L}}$ vs top mass for
the two methods.}
\label{d0mass}
\end{figure}

\begin{table}[h]
\begin{center}
\begin{tabular}{|c|c|c|c|c|}
\hline
Method & \mtop ($\gevcc$) & Fitted $N_s$ & Fitted $N_b$ \\ 
\hline\hline
LB     & 174.0$\pm$5.6       &  24$\pm$8    &   53$\pm$10  \\
NN     & 171.3$\pm$6.0       &  29$\pm$9    &   48$\pm$10  \\
LB+NN  & 173.3$\pm$5.6       &      -       &       -  \\
\hline
\end{tabular}
\caption{Fit results using the low-bias (LB) likelihood, then
neural-network (NN) method, and the combined result. Shown are
the fitted top quark masses, and the fitted number of signal ($N_s$)
and background ($N_b$) events.}
\label{t-d0mass}
\end{center}
\end{table}


\begin{table}[ht]
\begin{center}
\begin{tabular}{|c|c|c|}
\hline
Source 	    		&    CDF  & D0   \\
\hline\hline
Energy Scale   		&    4.4         &         4.0   \\
Signal Model 		&    1.8         &         1.9   \\
Background	   	&    1.3         &         2.5   \\
Noise/M.I.              &     -          &         1.3   \\
b-tagging bias		&    0.4         &          -   \\
Likelihood Method	&     -          &         1.3   \\   
\hline
Total			&    4.9         &         5.5  \\
\hline
\end{tabular}
\caption{Main systematic uncertainties ($\gevcc$) in the
l+jets top quark mass measurement for each experiment.
(M.I.=Multiple Interactions; PDF's = Parton Distribution Functions)}
\label{t-ljetsys}
\end{center}
\end{table}

\vspace{-0.15in}
\section{CDF All-Hadronic Top Mass Measurement}

   This measurement\cite{all-hadron} requires 6 or more jets
with $\Et>15 \ \gev$, and $|\eta |<2$, and at least one 
b-tagged jet. Additional cuts, based on MC simulations,
were employed to improve the S/B ratio. An event-by-event
top mass is evaluated using the measured jet energies and the
kinematic constraints mentioned previously. All thirty combinations
are tried, and the top mass having the lowest $\chi^2<10$ 
is taken as the top mass for the event.
The data sample consists of 136 events, of which 108$\pm$9 events
are expected to come from background sources. A likelihood 
procedure (with a gaussian background constraint) was used to 
extract a measured top mass  of 186$\pm$10(stat.)$\pm$12(syst.) $\gevcc$.
The systematic uncertainties are dominated by the jet energy scale 
signal modelling, and the fitting method.

\section{Combining the top mass measurements}

  The measurement in each decay topology may be combined to
improve the resolution on the top quark mass measurement. 
Each experiment combines their measurements into a single
measurement, taking into account correlations in uncertainties
between the measurements. Combining 
the three CDF measurements,
the top quark mass is measured to be 175.6$\pm$6.8 $\gevcc$.
The two D0 measurements yield a top quark mass of
172.1$\pm$7.1 $\gevcc$. These two mass measurements 
are combined assuming a 2 $\gevcc$ correlated systematic 
uncertainty. The
top quark mass at the Tevatron is therefore
measured to be 173.9$\pm$5.2 $\gevcc$. 
The values of $M_W$
versus \mtop for several (standard model) Higgs masses are shown in 
Fig.~3
The data tend to favor a mass 
below $\approx$500 $\gevcc$, but clearly 
%
more data is needed to reduce the 
uncertainties. In Run 2, we 
expect to measure \mtop to
within 1-2 $\gevcc$ and $M_W$ to within 40 ${\rm{MeV/c^2}}$.
This will clearly narrow the search window for future Higgs'
searches.

I would like to thank my collaborators at CDF for
their assistance in preparing this talk. I would also like to thank
the D0 top group, particularly Scott Snyder and Erich Varnes for
answering questions regarding their top mass analyses.

\begin{figure}[h]
\epsfxsize=6.0in
\hglue1truept\hspace{0.0in}
\epsffile{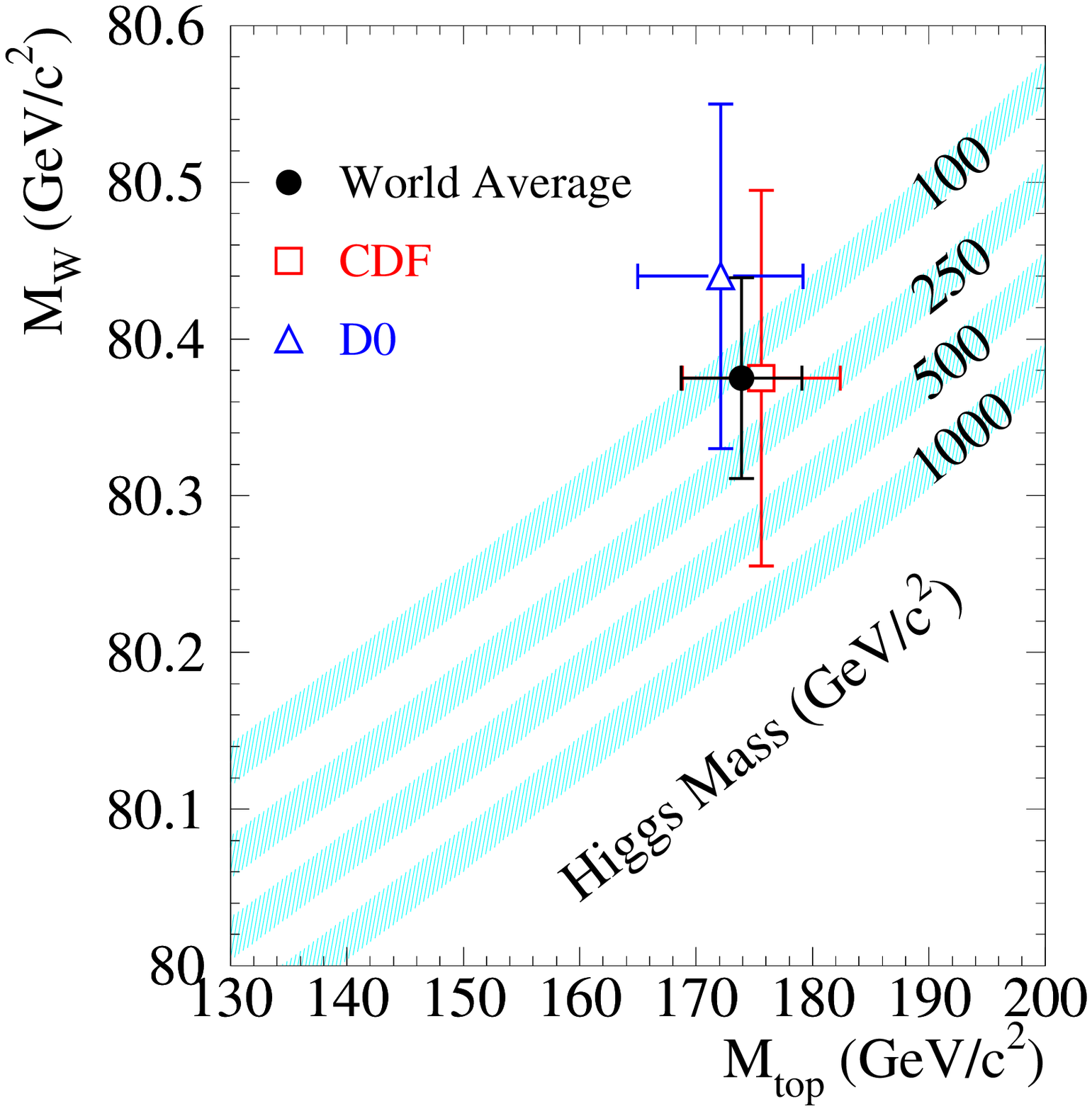}
\caption{$M_W$ versus \mtop for various choices of the Higgs mass.
Shown are the CDF and D0 averages, as well as the world average.}
\label{mwmt_cdf_d0_1998}
\end{figure}

\vskip1.2truein



\begin{thebibliography}{99}
\bibitem{metdef} The missing $\Et$, $\met$ is defined to be the magnitude of the 
vector sum of the transverse energy of all calorimeter towers.
\bibitem{bgfrac} The background fractions, $x_b$, are 
defined as the expected number of background events in a given subsample
divided by the total number of events in that subsample.
\end{thebibliography}
\end{document}